\documentclass[%
 reprint,
 amsmath,amssymb,
 aps,
]{revtex4-1}

\usepackage{graphicx}
\usepackage{dcolumn}
\usepackage{bm}

\usepackage{color}

\begin{document}

\title{Beyond the Ginzburg-Landau theory of freezing:\\Anisotropy of the interfacial free energy in the Phase-Field Crystal model}

\author{Gyula I. T\'oth}
\affiliation{Institute for Solid State Physics and Optics, Wigner Research Centre for Physics,\\P.O. Box 49, H-1525 Budapest, Hungary}
\email{toth.gyula@wigner.mta.hu}
\author{Nikolas Provatas}
\affiliation{Department of Physics and Centre for the Physics of Materials, McGill University, 3600 Rue University, Montreal, Canada H3A-2T8}

\date{\today}

\begin{abstract}
This paper re-visits the weakly fourth order anisotropic Ginzburg-Landau (GL) theory of freezing. First we determine the anisotropy of the interfacial free energy in the Phase-Field Crystal (PFC) model analytically, and prove that it remains finite at the critical point as a direct consequence of the one-mode dominance of the model. Next, we derive the leading order PFC amplitude model and show the formal analogy to traditional weakly 4th order anisotropic GL theories. We conclude that the material-independent anisotropy appearing in emergent GL theory coincides with the remnant anisotropy of the generating PFC model. As a result, we show that the reduced temperature $\epsilon$ does not enter into the interfacial free energy anisotropy for metallic materials in both the Phase-Field Crystal model and the emerging Ginzburg-Landau theories. Finally, we investigate the possible pathways of calibrating  anisotropic Ginzburg-Landau theories.

\end{abstract}

\maketitle


\section{Introduction}

The anisotropy of the crystal-liquid interfacial free energy is regarded as the key factor of dendritic solidification, since it determines the microstructure of the crystallizing material, including many commercial metallic alloys. Many attempts have been made to determine the shape and the value of the anisotropy of the interfacial free energy, including equilibrium shape measurements \cite{napolitano1,Liu20014271,PhysRevB.70.214103} and molecular dynamics simulations. Molecular dynamics-based methods, such as the cleaving technique \cite{:/content/aip/journal/jcp/84/10/10.1063/1.449884,PhysRevLett.85.4751,:/content/aip/journal/jcp/139/22/10.1063/1.4837695,PhysRevLett.94.086102} and the capillary fluctuation method \cite{PhysRevLett.86.5530,PhysRevB.66.144104} predict the anisotropy in the order of $1\%$ for several metallic systems. (For bcc systems, see References \cite{PhysRevB.69.020102,PhysRevB.69.174103}.) Since it has been revealed that the anisotropy critically depends on the crystal symmetry, and its magnitude depends mostly on the ratio of the crystal-liquid interface thickness and the interatomic distance, continuum descriptions also can be relevant tools for describing the anisotropic properties. 

The first order parameter theory that captures anisotropy was developed by Haymet and Oxtoby \cite{:/content/aip/journal/jcp/74/4/10.1063/1.441326,:/content/aip/journal/jcp/76/12/10.1063/1.443029}. The description is based on the classical Density Functional Theory (DFT) of freezing of the Ramakrishnan-Yussouff type \cite{PhysRevB.19.2775}, which chareacterizes the system by the time-averaged local one-particle density. Since the theory works on the molecular scale in space, it inherently contains the crystalline symmetries of the system. Later a more convenient description, the Ginzburg-Landau (GL) theory of bcc-liquid interfaces was developed by Shih et al. \cite{PhysRevA.35.2611}. In the GL theory the free energy of nonuniform phases is expressed in terms of space-dependent reciprocal lattice vector amplitudes, which are constant in the bulk phases and vary on the scale of the crystal-liquid interface thickness. The revised theory of Shih et al by Wu et al.  \cite{PhysRevB.73.094101} predicts $\nu \approx 3\%$ for iron. (In Reference \cite{PhysRevB.73.094101} the anisotropy parameter is defined as $\nu_{111}^{100}=(\gamma_{100}-\gamma_{111})/(\gamma_{100}+\gamma_{111})$, where $\gamma_{100}$ and $\gamma_{111}$ are the interfacial free energies for the $[100]$ and $[111]$ crystal-liquid equilibrium planar interfaces, respectively.) This value is also supported by the simpler, DFT motivated Phase-Field Crystal (PFC) model \cite{PhysRevLett.88.245701,doi:10.1080/00018732.2012.737555} and its amplitude theory \cite{PhysRevB.76.184107}, while two versions of the PFC model of Jaatinen et al \cite{PhysRevE.80.031602} yielded $\nu_{111}^{100}=3\%$ (GL-PFC) and $2.6\%$ (Eight-order fit PFC), respectively. 

Although the results of continuum theories are fair agreement with the experimental results and the results of atomistic simulations, both the $4^{th}$-order GL and PFC amplitude theories of pure materials have a quite worrisome common property pointed out by Majaniemi and Provatas \cite{PhysRevE.79.011607}: they are "weak" in a manner that all material parameters (except the crystal structure) scale out from the free energy functional. Consequently, the anisotropy parameters in these models depend exclusively on the crystal structure but not on the temperature, which results in a limited applicability of these models, and necessitates proper modifications. Such modifications may be including further reciprocal lattice vector sets and/or applying higher order polynomials in the free energy density \cite{PhysRevA.35.2611}.

The starting point of developing consistent anisotropic Ginzburg-Landau theories is classical Density Functional Theory. The classical DFT inherently contains the crystal symmetries, and its amplitude expansions lead to particular Ginzburg-Landau theories. Since the PFC is a $4^{th}$-order density functional theory with relatively simple spatial operators, it is a good candidate  to employ for showing the relationship between the mathematical form of the anisotropy in a GL theory and how it emerges from an underlying classical DFT. In addition, the PFC amplitude theories show formal analogy to the anisotropic GL theories in the sense of the "weak" nature, which seems to be more than just a coincidence.

The paper is organized as follows: In Section II we discuss the invariant formulations of the Phase-Field Crystal free energy functional. In Section III we calculate the equilibrium properties of the bulk (liquid and crystal) phases, and determine the properties (exponents and coefficients) of the equilibrium crystal amplitude and equilibrium density. Using the results, in Section IV we calculate the interfacial free energy, and prove that the anisotropy remains finite at the critical point, which is a direct consequence of the one-mode dominant behavior of the PFC. Finally, we derive the free energy functional of the anisotropic amplitude expansion of the PFC model in the leading order, and show that it is equivalent to a weakly fourth order Ginzburg-Landau theory. In section V we discuss the results.

\section{The Phase-Field Crystal model}

In the first part we investigate the crystal-liquid equilibrium in the Phase-Field Crystal model introduced in Ref. \cite{PhysRevLett.88.245701}. After defining the free energy functional, we investigate the behavior of the PFC model close to the critical point, and prove that the first reciprocal lattice vector (RLV) set dominance of the model is related to the critical exponents of the RLV set amplitudes.

\begin{figure}
\includegraphics[width=0.49\linewidth]{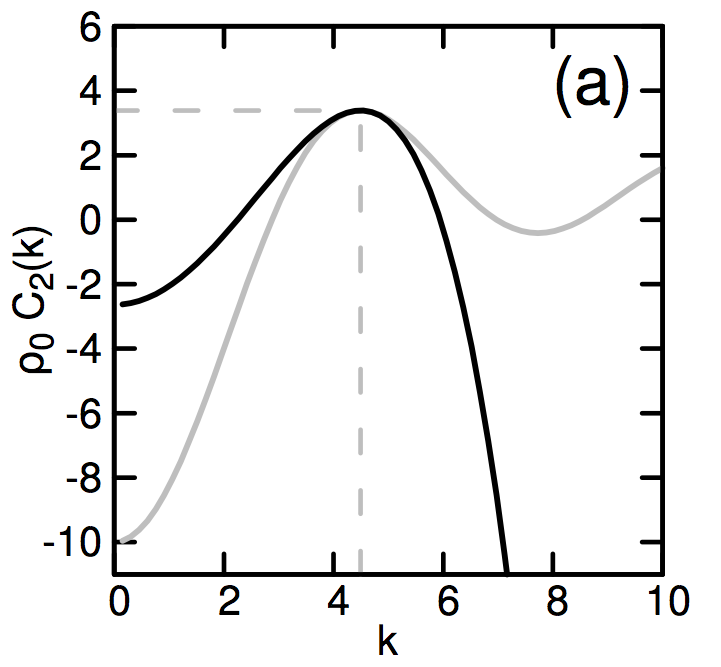}
\includegraphics[width=0.48\linewidth]{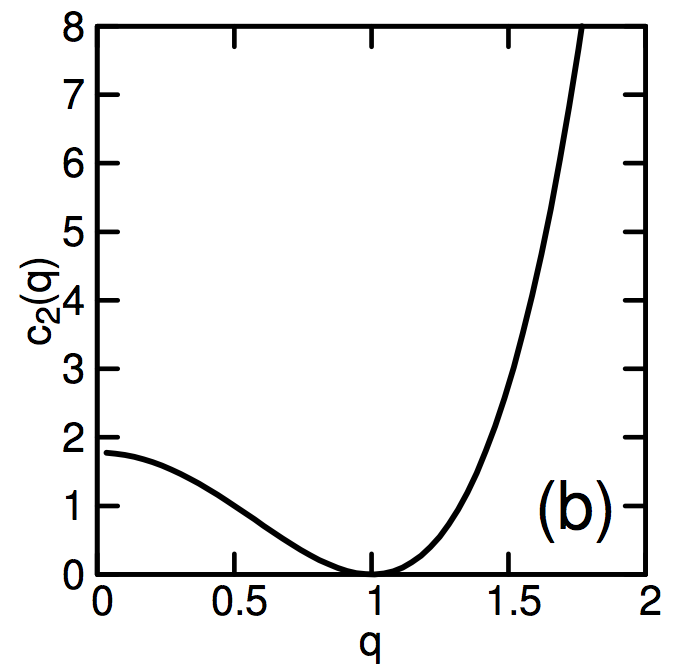}
\caption{Direct correlation functions: (a) Schematic correlation function of a real system (gray) and typical Phase-Field Crystal correlation function (black). (b) Scaled PFC correlation function $c_2(q)=1-C_2(q \cdot k_0)/C_2(k_0)$, where $k_0$ is the position of the maximum of the PFC $C_2(k)$ (indicated by the horizontal dashed gray line in panel a). Note that the zero-valued minimum of $c_2(q)$ at $q=1$ is independent from the particular form of the PFC $C_2(k)$.}
\end{figure}

\subsection{Minimal form of the free energy functional}

In the single-component Phase-Field Crystal model the free energy of the system relative to a reference homogeneous state of density $\rho_0$ reads as \cite{PhysRevLett.88.245701}:
\begin{equation}
\label{eq:model_general}
\frac{\Delta F}{\rho_0 k_B T}= \int d\mathbf{r} \left\{ n \frac{1-\rho_0\hat{C}_2}{2} n - a \frac{n^3}{3} + b \frac{n^4}{4} \right\} \enskip ,
\end{equation}
where $n(\mathbf{r})=[\rho(\mathbf{r})-\rho_0]/\rho_0$ is the scaled density field, and $C_2(k)$ is a single-peaked direct correlation function in the wavelength space with peak position $k_0$ (see Fig 1.a). As a first step, we scale the model in order to identify the important parameters: Scaling the length as $\mathbf{r}=\lambda \cdot \tilde{\mathbf{r}}$, the order parameter as $n=X \cdot \phi$ and the free energy as $\Delta F/(\rho_0 k_B T)=A \cdot \tilde{F}$ results in a simplified form of Eq. (\ref{eq:model_general}):
\begin{equation}
\label{eq:model_base0}
\tilde{F}= \int d\tilde{\mathbf{r}} \left\{ \phi \frac{\hat{{c}}_2-r}{2} \phi  - t \frac{\phi^3}{3} + \frac{\phi^4}{4} \right\} \enskip .
\end{equation}
The choice of $\lambda:=1/k_0$ and
\begin{equation}
\label{eq:c2transform}
c_2(q):=[C_2(k_0)-C_2(q \cdot k_0)]/v
\end{equation}
results in the scales $X=\sqrt{\rho_0 v/b}$ and $A=[\rho_0 v]^2/(k_0^3 b)$, and the parameters $$r=\frac{\rho_0 C_2(k_0)-1}{\rho_0 v}$$ and $t=a/\sqrt{b \rho_0 v}$. Here $v > 0$ is an arbitrary scaling parameter: for example, choosing $v=C_2(k_0)-C_2(0)$ generates $c_2(0)=1$. Note that $c_2(q)$ is a non-negative function with a single minimum at $q_0=1$ with $c_2(1) = 0$ (see Fig 1.b). This transformation of the direct correlation function will play a crucial role in our derivation. Taking into account that $c_2(q)$ is an even function, it can be written as $c_2(q)=\sum_{i=0}^\infty \alpha_i q^{2i}$, which corresponds to $\hat{c}_2=\sum_{i=0}^\infty \alpha_i (-\nabla^2)^i$ in real space. (For the sake of simplicity, we won't use $\tilde{.}$ from this point). Consequently, the term $\phi \, \hat{c}_2 \, \phi=\sum_{i=0}^\infty \alpha_i (-1)^i \phi [ \nabla^{2i} \phi]$ in Eq. (\ref{eq:model_base0}) is equivalent to $\sum_{i=0}^\infty \alpha_i (\nabla^i \phi) ^2$ in the variational sense (note that both formulae results in the same functional derivative with respect to $\phi$). Using this equivalence, the cubic term $-t(\phi^3/3)$ can be eliminated: Substituting $\phi=\psi+t/3$ into Eq. (\ref{eq:model_base0}) simply results in $\phi \, \hat{c}_2 \, \phi \to \psi \, \hat{c}_2 \, \psi$, while the terms \textit{up to the first order} in $\psi$ can be neglected (since such terms vanish in both the Euler-Lagrange equation and the equation of motion). The "minimal" form of the original free energy functional then reads as
\begin{equation}
\label{eq:model_base}
F = \int d\mathbf{r} \left\{ \psi \frac{\hat{{c}}_2-\epsilon}{2} \psi  + \frac{\psi^4}{4} \right\} \enskip ,
\end{equation}
where $\epsilon=r-t^2/3$. This is a fairly simple form compared to Eq. (\ref{eq:model_general}) and shows that the important parameters of the model are only $\epsilon$ and $c_2(q)$.

\subsection{Periodic solutions}

Eq. (\ref{eq:model_base}) generates a first order phase transition between homogeneous (liquid) and lattice periodic (crystal) solutions. These phases represent extrema of the free energy functional, therefore, they can be found by solving the Euler-Lagrange equation: $\delta F/\delta \psi=\mu$ by definition, where $\delta F/\delta\psi$ is the functional derivative of $F$ with respect to $\psi$, and $\mu=(\delta F/\delta \psi)_{\psi_L}$, i.e. the chemical potential of a homogeneous background liquid of density $\rho_L$.  Since the ELE is a nonlinear, higher order PDE,  usually it is solved numerically. Instead, however, we can parametrize the lattice periodic solution in the following general form: 
\begin{equation}
\label{eq:lpfunc}
\psi_p(\mathbf{r}) = \bar{\psi} + \sum_I A_I \sum_{i \in S(I)} \exp^{\imath \mathbf{\Gamma}_i^I \cdot \mathbf{r}} \enskip ,
\end{equation}
where $\bar{\psi}$ is the average density, $A_I$ the amplitude of the $I^{th}$ RLV set, and $\mathbf{\Gamma}_i^I$ the $i^{th}$ RLV in the $I^{th}$ RLV  set. The \textit{bulk} free energy density is defined as the volumetric average of the free energy in a unit cell:
\begin{equation}
\label{eq:fdensdef}
f[\psi_p] := \frac{1}{V_{cell}}\int_{V_{cell}} dV \{ I[\psi_p] \} \enskip ,
\end{equation}
where $I[.]$ is the integrand of Eq. (\ref{eq:model_base}). For practical reasons we define the free energy density difference as:
\begin{equation}
\label{eq:dfdensdef}
\Delta f[\psi_p] := f[\psi_p]-f[\bar{\psi}] \enskip .
\end{equation}
Using the definitions (\ref{eq:fdensdef}) and (\ref{eq:dfdensdef}), and substituting Eq. (\ref{eq:lpfunc}) into Eq. (\ref{eq:model_base}) together with $\psi \cdot \hat{c}_2[\psi]=\sum_{i=0}^\infty \alpha_i (\nabla^i \psi)^2$ results in (see Appendix A):
\begin{equation}
\begin{split}
\label{eq:dfbase}
\Delta f[\psi_p] = &\sum_{I} \left[ A_I^2 N^{(2)}_{I,I} \right] \frac{c_2(\Gamma_I) - \epsilon + 3\bar{\psi}^2}{2} + \\
& + \bar{\psi} \sum_{I,J,K} (A_I A_J A_K) \mathcal{N}^{(3)}_{I,J,K} \\
& + \frac{1}{4}\sum_{I,J,K,L} (A_I A_J A_K A_L) \mathcal{N}^{(4)}_{I,J,K,L} \enskip ,
\end{split}
\end{equation}
where we introduced the shorthand notation
\begin{equation}
\label{eq:countingop}
\mathcal{N}^{(N)}_{I_1,I_2, \dots, I_N} := \sum_{i_1,i_2,\dots,i_N} \delta_{i_1,i_2,\dots,i_N}^{I_1,I_2,\dots,I_N} \enskip ,
\end{equation}
where  $\delta_{i_1,i_2,\dots,i_N}^{I_1,I_2,\dots,I_N}$ denoted here as the Kronecker-delta function $\delta(\mathbf{\Gamma}_{i_1}^{I_1}+\mathbf{\Gamma}_{i_2}^{I_2}+\dots+\mathbf{\Gamma}_{i_N}^{I_N})$, which gives 1 if the sum of the reciprocal lattice vectors in the argument is zero, otherwise it is 0. Therefore, $\mathcal{N}^{(N)}_{I_1,I_2, \dots, I_N} $ is just the total number of N-term vector sums resulting in zero in which the first vector is from the RLV set $I_1$, the second is from $I_2$ and so on. Consequently, $N^{(2)}_{I,I}$ is just the number of RLVs in the $I^{th}$ RLV set. Note that $\mathcal{N}^{(N)}_{I_1,I_2, \dots ,I_N}$ is invariant for the permutation of the indices.
 
\subsection{Equilibrium conditions}

Eq. (\ref{eq:dfbase}) realizes a parametrization of the free energy functional, which has to be minimized with respect to the set amplitudes $A_I$ and the selected wavelength $\Gamma_I$ at a constant average density $\bar{\psi}$. Introducing $\Gamma_I=\beta_I q$, where $\beta_1=1$, the minimization equations read as:
\begin{equation}
\label{eq:mini}
\frac{\partial \Delta f[\psi_p]}{\partial A_I} = 0 \quad \text{and} \quad \frac{\partial \Delta f[\psi_p]}{\partial q} = 0 \enskip .
\end{equation}
From Eq. (\ref{eq:mini}) two qualitatively different types of solutions emerge: (i) the trivial solution: $A_I \equiv 0$ for $I=1\dots\infty$ (homogeneous solution, the liquid phase), and (ii) a nontrivial lattice periodic solution (crystalline phase), where $A_I \neq 0$. Neglecting the crystal-liquid density jump for $0<\epsilon\ll 1$ the crystal-liquid equilibrium is simply defined by equal free energy densities of the phases at the same average density, i.e. 
\begin{equation}
\label{eq:eqcond}
f[\bar{\psi}] = f[\psi_p] \quad \Rightarrow \quad \Delta f[\psi_p]=0 \enskip ,
\end{equation}
where $\psi_p$ is the nontrivial solution. Eq. (\ref{eq:eqcond}) together with Eq. (\ref{eq:mini}) defines the atomic distance $q$, the equilibrium solid amplitudes $A_I$ and the equilibrium density $\bar{\psi}$ as a function of $\epsilon$ and $c_2(q)$.

\subsection{Critical behavior}

In this section we show that the general PFC model described by Eq. (\ref{eq:model_general}) generates a mean-field Brazowskii/Swift-Hohenberg \textit{critical point} at $\epsilon =0$. We determine the critical exponents of the equilibrium density ($y_\psi$) and crystal RLV set amplitudes ($y_I$) and show that $y_1<y_I$ for any $I>1$, implying the the one-mode dominance of the model.

\subsubsection{Wavelength selection}

For the particular choice $c_2(q)=(1-q^2)^2$ Eq. (\ref{eq:model_base}) reduces to the well-known Brazowskii/Swift-Hohenberg form, which has a critical point at $\epsilon=0$ \cite{PhysRevE.70.051605}. It is reasonable to assume that this behavior doesn't depend on the particular form of $c_2(q)$, and the model has a critical point as long as $c_2(q)$ is a positive semidefinite function with a single, zero-value minimum at $k=1$, i.e. $c_2(1)=0$. Indeed, it is relatively easy to see that the only solution of Eqns. (\ref{eq:mini}) and (\ref{eq:eqcond}) for $\epsilon=0$ is $\bar{\psi}=0$ and $A_I=0$. Therefore, we can write $A_I = a_I \epsilon^{y_I} + h.o.t.$ and $\bar{\psi} = c_\psi \epsilon^{y_\psi} + h.o.t.$ for $0<\epsilon\ll 1$ in general. In order to determine the critical exponents first we \textit{assume} that there are more than one dominant RLV sets, meaning that  $y_{I_1}=y_{I_2}=\dots=y_{I_N}(=:y_A)$ , where $N>1$ and $y_J > y_A$ for all $J \neq I_1,I_2,\dots I_N$. Using this, the leading order term of Eq. (\ref{eq:eqcond}) reads as:
\begin{equation}
\label{eq:lead}
\sum_{I \in \{ I_1,I_2,\dots,I_N \} } a_I^2 \mathcal{N}^{(2)}_{I,I} c_2(\beta_I q_0) = 0 \enskip ,
\end{equation}
where $q_0$ is the selected wavelength satisfying $(\partial \Delta f[\psi_p]/\partial q_0)|_{q_0}=0$. Since $a_I^2 \mathcal{N}^{(2)}_{I,I} >0$ and $c_2(q) \geq 0$, Eq. (\ref{eq:lead}) can be satisfied only if $c_2(\beta_I q)=0$ for \textit{all} dominant RLV sets. Since $c_2(q)$ has only one minimum at $q_0=1$ for which $c_2(1)=0$, only one RLV set can be dominant. In addition, this must be the first RLV set (thus $q=q_0$), since we're searching for a crystal structure (in other words, the only dominant RLV set cannot be a harmonic). Moreover, since $c_2(1)=0$, \textit{the term $\psi \, \hat{c}_2 \,\psi$ has no effect on the phase diagram}. This is in accordance with the original assumption, that the existence of the critical point doesn't depend on the particular choice of $c_2(q)$. The critical point exists as long as $c_2(q) \geq 0$ and has a single minimum at $q_0=1$ with $c_2(q_0)=0$.

\subsubsection{Critical exponents}

Taking into account that $y_I>y_A$ for $I>1$ and using $q_0=1$, the equilibrium condition reads as:
\begin{equation}
\label{eq:expand}
\begin{split}
& \Delta f[\psi_p] = A_1^2 N_1 \frac{3\bar{\psi}^2-\epsilon}{2} + \bar{\psi} A_1^3 N_3 + \frac{A_1^4}{4} N_4 + \\
&+ \sum_{I>1} \left[ A_I^2  N_I c_2(\beta_I) + \right. \\
& \left. + 3 \bar{\psi} A_1^2  A_{I} \mathcal{N}^{(3)}_{1,1,I}  +  A_1^3 A_{I} \mathcal{N}^{(4)}_{1,1,1,I}  + h.o.t. \right] = 0\enskip ,
\end{split}
\end{equation} 
where we used the shorthand notations $N_I:=\mathcal{N}_{I,I}^{(2)}$, $N_3:=\mathcal{N}_{1,1,1}^{(3)}$ and $N_4:=\mathcal{N}_{1,1,1,1}^{(4)}$ (details are shown in Appendix A). From Eq. (\ref{eq:lpfunc}) it is trivial that $y_A=y_\psi$, otherwise, there is no first order transition for $\epsilon \to 0$. In addition, in order to find nontrivial solution for $a_1$ and $c_\psi$, the $\propto\psi^4$ term in the free energy functional must contribute to the leading order. Taking these facts into account, the first row of Eq. (\ref{eq:expand})  together with $\partial \Delta f[\psi_p]/\partial A_I=0$ implies
\begin{equation}
\label{eq:critical_exponents}
y_A=y_\psi=1/2 \enskip ,
\end{equation}
therefore, the leading order of Eq. (\ref{eq:dfbase}) is $\epsilon^2$. In the next order of Eq. (\ref{eq:expand}) (the second and the third lines) the minimization equations for $A_{I>1}$ are decoupled:
\begin{equation}
\begin{split}
&\frac{\partial \Delta f[\psi_p]}{\partial A_I} =  2 A_I N_I c_2(\beta_I) + \\
& + 3A_1^2 \bar{\psi} \mathcal{N}^{(3)}_{1,1,I} +  4 A_1^3 \mathcal{N}^{(4)}_{1,1,1,I} + h.o.t. = 0 \enskip ,
\end{split}
\end{equation}
resulting in
\begin{equation}
y_{I>1}=3/2
\end{equation}
on the same basis, therefore, the next order of Eq. (\ref{eq:expand}) is proportional to $\epsilon^3$. In addition, from $\partial \Delta f[\psi_p]/\partial q=0$ it can be shown that $q^2=1+O(\epsilon^2)$, therefore, the first correction from this in Eq. (\ref{eq:expand}) is in the order of $\epsilon^4$. This means that our calculation is self-consistent.

Finally, one can determine the coefficients $c_\psi$ and $a_1$ by substituting $q_0=1$, $A_1=a_1\sqrt{\epsilon}$, $A_{I>1}=a_I \epsilon^{3/2}$ and $\bar{\psi}=c_\psi \sqrt{\epsilon}$ into Eq. (\ref{eq:dfbase}) then taking the leading order of Eqns. (\ref{eq:mini}) and (\ref{eq:eqcond}). The equations then can be solved analytically for $c_\psi$ and $a_1$:
\begin{eqnarray}
\label{eq:coefficient1}
c_\psi &=& - \sqrt{\frac{N_1 N_4}{3 N_1 N_4 - 2 N_3^2}} \enskip , \\
\label{eq:coefficient2}
a_1 &=& \sqrt{\frac{4 N_1 N_3^2}{N_4(3 N_1 N_4-2 N_3^2)}} \enskip ,
\end{eqnarray}
showing that the leading order equilibrium density and crystal amplitude depend exclusively on the crystal structure (apart from $\sqrt{\epsilon}$, naturally).\\

It is noteworthy that our results stay valid when the equilibrium density jump is considered in the calculations (for details, see Appendix B).

\section{Interfacial free energy}

In this section first we define the crystal-liquid interfacial free energy in the Phase-Field Crystal model, then we will approximate it analytically by using the results of the previous section. Considering the isotropic case first, we determine the interface thickness(es) and the interfacial free energy, and their critical exponents. As a key contribution of this work, we prove that the one-mode dominance of the PFC model, shown in the previous section, results in a remnant equilibrium crystal-liquid interfacial free energy anisotropy at the critical point. In the final part of this section we will determine the remnant anisotropy for the bcc structure and verify the result by comparing it to the results of numerical solutions of the Euler-Lagrange equation.

\subsection{Definition of the anisotropic crystal-liquid interfacial free energy}

When the density jump between the equilibrium crystal and liquid is neglected, the anisotropic interfacial free energy reads as 
\begin{equation}
\label{eq:gamma_base}
\gamma(\mathbf{n}) = \int_{-\infty}^{\infty} d\xi \left( \frac{1}{A_\perp}\int_\xi dA_\perp \left\{ \Delta I[\psi_{sl}] \right\} \right) \enskip ,
\end{equation} 
where $\mathbf{n}$ is the normal of the planar crystal-liquid interface, $\xi=\mathbf{n}\cdot\mathbf{r}$ the orthogonal distance from the interface, while $(1/A_\perp)\int_\xi dA_\perp\{.\}$ denotes an average calculated for a plane parallel to the interface at a constant value of $\xi$. The integrand of Eq. (\ref{eq:gamma_base}) reads as
\begin{equation*}
\Delta I[\psi_{sl}(\mathbf{r})]= I[\psi_{sl}(\mathbf{r})]-I[\bar{\psi}] \enskip .
\end{equation*}
Here $\psi_{sl}(\mathbf{r})=\bar{\psi}+\Delta \psi_{sl}(\mathbf{r})$ represents the equilibrium crystal-liquid density distribution, where $\Delta \psi_{sl}(\mathbf{r})$ is approximated as
\begin{equation}
\label{eq:slprof}
\Delta \psi_{sl}(\mathbf{r}) \approx \sum_I A_I \sum_{i \in S(I)} \frac{1+g_i^I(\xi)}{2} h_i^I(\mathbf{r})  \enskip ,
\end{equation}
where $A_I=a_I \epsilon^{y_I}$ (the equilibrium crystal amplitudes) and we have used the following shorthand notations:
\begin{equation}
\label{eq:subsgh}
g_i^I(\xi) := \tanh(\xi/\Lambda_i^I) \quad \text{and} \quad h_i^I(\mathbf{r}) := \exp(\imath \mathbf{\Gamma}_i^I \cdot \mathbf{r}) \enskip ,
\end{equation}
where $\Lambda_i^I$ is the characteristic interface width of the $i^{th}$ plane wave in the $I^{th}$ RLV set \cite{PhysRevE.79.011607}. Note that far from the interface Eq. (\ref{eq:slprof}) recovers the density distribution of the equilibrium bulk phases: $\psi_{sl}(\mathbf{r})|_{\xi\to+\infty} \to \psi_p(\mathbf{r})$ and $\psi_{sl}(\mathbf{r})|_{\xi\to-\infty} \to \bar{\psi}$. Using Eqns. (\ref{eq:slprof}) and (\ref{eq:lpfunc}) in Eq. (\ref{eq:tension_definition}), after a straightforward but lengthy algebra one can come to a reasonably simple parametrized form of the \textit{leading order} anisotropic crystal-liquid interfacial free energy (for details, see Appendix C):
\begin{equation}
\label{eq:gamma_simp}
\begin{split}
&\gamma (\mathbf{n}) = A_1^2 \sum_{i} \left[ \frac{\epsilon-3\bar{\psi}^2}{4} \Lambda_i^1 + \frac{2\,\zeta(\mathbf{n} \cdot \mathbf{\Gamma}_i^1)}{3\,\Lambda_i^1}\right]  + \\
& \frac{A_1^3\bar{\psi}}{8} \sum_{i,j,k} \left[ \sum_{m,n}^{i,j,k} \|g_m^1 g_n^1 -1 \| \right] \delta_{i,j,k}^{1,1,1} + \\
& \frac{A_1^4}{64} \sum_{i,j,k,l} \left[ \| g_i^1 g_j^1 g_k^1 g_l^1 -1 \| + \sum_{m,n}^{i,j,k,l} \| g_m^1 g_n^1 -1\| \right] \delta_{i,j,k,l}^{1,1,1,1} \enskip ,
\end{split}
\end{equation}
where the sums for $(m,n)$ run for all \textit{different} pairs in $(i,j,k)$ and $(i,j,k,l)$, respectively, while we used the shorthand notation $\| . \|:=\int_{-\infty}^{+\infty} d\xi \{.\}$. The function
\begin{equation}
\zeta(x)=\zeta_0 + \zeta_1 \cdot x^2
\end{equation}
is responsible for the anisotropic contribution [here $\zeta_0$ and $\zeta_1$ are constants emerging from the particular form of $c_2(q)$]. For example, for the $c_2(q)=(1-q^2)^2$ theory (Brazowskii/Swift-Hohenberg), $\zeta(\mathbf{n} \cdot \mathbf{\Gamma}_i^1)=(\mathbf{n} \cdot \mathbf{\Gamma}_i^1)^2$. Note, that the appearance of the anisotropic contribution to the leading order of $\gamma(\mathbf{n})$ is the consequence of the one-mode dominance of the theory, i.e. $y_1 < y_I$ for any $I>1$.\\

\subsection{Critical exponent of the interface thickness}

Close to the critical point the interface thickness (correlation length) diverge as $\Lambda_i^I = \lambda_i^I  \cdot \epsilon^{y_\Lambda}$, where $y_\Lambda<0$. Note that all interface thicknesses diverge with the unique critical exponent $y_\Lambda$ (for details, see Appendix D). In case of the isotropic limit ($\Lambda_i^1=\Lambda_i^2=\dots=\Lambda$), Eq. (\ref{eq:gamma_simp}) reads as:
 \begin{equation}
\label{eq:gamma_power}
\begin{split}
\gamma_{\rm iso}  = & A_1^2 N_1 \left(  \frac{\epsilon-3\bar{\psi}^2}{4} \Lambda + \frac{2 C}{3\Lambda} \right) - \\
& - \left[ \frac{3\bar{\psi}}{4} A_1^3 N_3 + \frac{11}{48} A_1^4 N_4 \right] \Lambda \enskip ,
 \end{split}
\end{equation}
 where
 \begin{equation}
 C=\frac{1}{N_1}\sum_{i \in S(1)}\zeta(\mathbf{n}\cdot \mathbf{\Gamma}_i^1)
 \end{equation}
 is constant for geometrical reasons, and we used that $\|(g_i^1)^2-1\|=-2\Lambda$ and $\|(g_i^1)^4-1\|=-(8/3)\Lambda$. Using $\Lambda = \lambda \cdot \epsilon^{y_\lambda}$ (where $\lambda$ is a constant specific to the isotropic case) in the minimization equation $\partial \gamma/\partial \Lambda=0$ yields
 \begin{equation}
 y_\Lambda = -1/2 \enskip ,
 \end{equation}
and
\begin{equation}
\label{eq:isolambda}
\frac{1}{\lambda^2} =  \left(\frac{1}{8C}\right) \frac{N_3^2}{3 N_1 N_4 - 2 N_3^2} \enskip .
\end{equation}
Using these in Eq. (\ref{eq:gamma_power}) the isotropic interfacial free energy reads as
\begin{equation}
\label{eq:isogamma}
\frac{\gamma_{\rm iso}}{\epsilon^{3/2}} = \frac{4}{3} \frac{N_1^2}{N_4}\sqrt{2 C} \left( \frac{3 N_1 N_4- 2 N_3^2}{N_3^2} \right)^{-3/2} \enskip .
 \end{equation}
 Note that the particular form of $\hat{c}_2$ appear exclusively in the constant $C$. Moreover, $C$ scales as $C \to C/v$ with $v$ from Eq. (\ref{eq:c2transform}), and $\epsilon$ as $\epsilon \to \epsilon/v$ (since $\epsilon=r-t^2/3 \propto 1/v$), which results in the simple scaling relation
 \begin{equation}
 \label{eq:gammascaled}
 \gamma_{v}(\epsilon)/v= \gamma_1(\epsilon/v) \enskip ,
 \end{equation}
 where $\gamma_1$ and $\gamma_v$ denote the isotropic interfacial free energy at $v=1$ and an arbitrary $v$, respectively. Eq. (\ref{eq:gammascaled}) shows that $v$ is not a relevant parameter of the theory, and only helps to choose a convenient form of $c_2(q)$.
 
\subsection{Critical behavior of the anisotropy}

Using the critical exponents and the facts that $\|g_i^1 g_j^1-1\| \propto \epsilon^{y_\Lambda}$ and $\|g_i^1 g_j^1 g_k^1 g_l^1 -1\| \propto \epsilon^{y_\Lambda}$ in Eq. (\ref{eq:gamma_simp}) yields
\begin{equation}
\label{eq:gamma_factorized}
\frac{\gamma(\mathbf{n})}{\epsilon^{3/2}} = (c^{i}_0+ c^{i}_1 \epsilon + \dots) + [c^{a}_0(\mathbf{n}) + \epsilon^2 c^{a}_2(\mathbf{n}) + \dots ] \enskip ,
\end{equation}
where the indices $()^{i,a}$ denote isotropic and anisotropic contributions, respectively. The anisotropy parameter is defined as
\begin{equation}
\label{eq:anisotropy}
\nu_{\min}^{\max} := \frac{\max[\gamma(\mathbf{n})]-\min[\gamma(\mathbf{n})]}{\max[\gamma(\mathbf{n})]+\min[\gamma(\mathbf{n})]} \enskip .
\end{equation}
Applying Eq. (\ref{eq:gamma_factorized}) in Eq. (\ref{eq:anisotropy}) results in
\begin{equation}
\nu_{\min}^{\max}(\epsilon) = \frac{[c_0^a(\mathbf{n}^+)-c_0^a(\mathbf{n}^-)] + O(\epsilon^2)}{2 c_0^i+c_0^a(\mathbf{n}^+)+c_0^a(\mathbf{n}^-)+O(\epsilon)} \enskip ,
\end{equation}
where $\mathbf{n}^\pm$ are defined by $\gamma(\mathbf{n}^+):=\max[\gamma(\mathbf{n})]$ and $\gamma(\mathbf{n}^-):=\min[\gamma(\mathbf{n})]$, respectively. From Eq. (\ref{eq:gamma_simp}) one can see that $c_0^a(\mathbf{n}^\pm) \propto \sum_i \zeta(\mathbf{n}^\pm \cdot \mathbf{\Gamma}_i^1)/\lambda_i^1$. However, $\lambda^1_i \neq \lambda^1_j$ for $i \neq j$  in case of $\zeta_1 \neq 0$ in $\zeta(x)$, therefore, \textit{the anisotropy is seen to be finite at the critical point}:
\begin{equation}
\label{eq:remnant}
\nu_{\min}^{\max}(\epsilon) = \nu_0 + O(\epsilon) \enskip ,
\end{equation}
which apparently contradicts to former expectations of Podmaniczky et al. \cite{Podmaniczky2014148}. Note that the remnant anisotropy ($\nu_0$) is a direct consequence of the one-mode dominant nature of the free energy functional: $y_1=1/2$ together with $y_{I>1}=3/2$ may yield a non-vanishing anisotropic contribution to the leading order of to the interfacial free energy.\\

\begin{figure}
\includegraphics[width=1.0\linewidth]{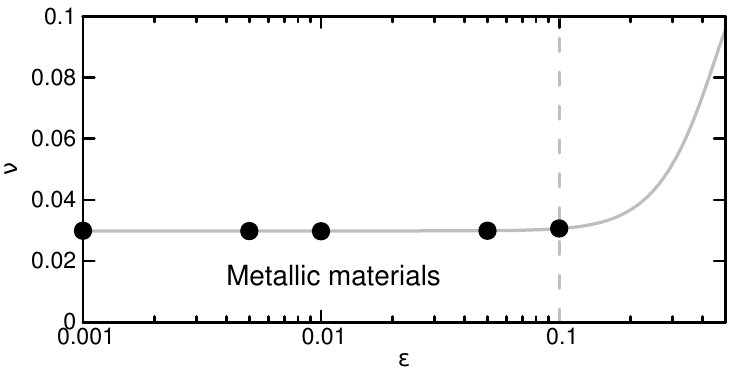}
\caption{Crystal-liquid interfacial free energy anisotropy for the $c_2(q)=(1-q^2)^2$ (Brazowskii/Swift-Hohenberg) model. For $\epsilon<0.1$, $\nu_{\min}^{\max}=\nu_{111}^{100}$.}
\end{figure}

We have to mention at this point that it would also be useful to investigate the critical behavior of the anisotropy in the presence of fluctuations. Fluctuations destroy the mean-field behavior, and we know that the anisotropy vanishes at the critical point in the triangular Ising system \cite{PhysRevB.63.085410}, however, we also know that the Brazowskii system has its own universality class \cite{e_041_01_0085,JChemPhys.91.7265}. 

\subsection{Verification of the remnant anisotropy}

To determine $\nu_0$ in Eq. (\ref{eq:remnant}), first we calculate Eq. (\ref{eq:gamma_simp}) divided by Eq. (\ref{eq:isogamma}) for the general anisotropic case using a reasonable approximation of the envelope function integrals, which are defined as $\|g_i^1 g_j^1-1\|$ and $\|g_i^1 g_j^1 g_k^1 g_l^1-1\|$ in Eq. (\ref{eq:gamma_simp}) and approximated in detail Appendix D. It yields the coupled minimization equations for interface thickness constants relative to the isotropic one, i.e. $\tilde{\lambda}_i = \Lambda_i^1/\Lambda_1$:
\begin{equation}
\frac{\partial}{\partial \tilde{\lambda}_i}\left[ \frac{\gamma(\mathbf{n})}{\gamma_{\rm iso}}\right] = 0 \quad  (i=1\dots N_1) \enskip ,
\end{equation}
which have to be solved numerically for $\tilde{\lambda}_i$ for the $c_2(q)=(1-q^2)^2$ model. For the bcc structure preferred by the Phase-Field Crystal model close to the critical point in 3 dimensions $N_1=12$, $N_3=48$ and $N_4=480$. We started the numerical calculations from the isotropic solution defined by Eq. (\ref{eq:isolambda}) for the [111] and [100] crystal planes, which give the minimal and the maximal interfacial free energies, respectively. Our calculation resulted in a significant remnant anisotropy
\begin{equation}
\nu_0 = (2.6 \pm 0.01)\% \enskip .
\end{equation}
For comparison, following the method of Podmaniczky et al. \cite{Podmaniczky2014148} we evaluated the interfacial free energy by solving the Euler-Lagrange equation $\delta F/\delta \psi=\mu$ numerically for bcc-liquid equilibrium interfaces at $\epsilon=0.001,0.005,0.01,0.05$ and $0.1$. We have found $\nu_0(\epsilon)=(3\pm0.05)\%$, a nearly constant anisotropy parameter, which is in a  fair agreement with the analytical result, and moreover, is in a perfect agreement with the results from the GL theory of Wu et al. \cite{PhysRevB.73.094101} or the PFC amplitude equations of Wu and Karma \cite{PhysRevB.76.184107}. This unexpected coincidence, however, suggests a deeper relationship between the weakly $4^{th}$ order Ginzburg-Landau / amplitude models of classical density functional theories having a critical point.

\section{Connection to Ginzburg-Landau theories}

In this section we will investigate the connection between the critical behavior of the Phase-Field Crystal model and Ginzburg-Landau theories. First we derive an isotropic amplitude model from the critical PFC model (i.e. in the leading order in case of $\epsilon \to 0$), then - following the recent work of Provatas and Majaniemi - extend it for the anisotropic case. A key result of this paper is to formally show that the leading-order amplitude model of the PFC close to the critical point is analogous to a weakly $4^{th}$ order anisotropic one-mode Ginzburg-Landau theory, and the material parameter independent interfacial free energy anisotropy appearing in the GL theory is precisely the critical point remnant anisotropy inherited from the generating density functional theory.

\subsection{Isotropic limit}

\subsubsection{Ginzburg-Landau polynomial}

In equilibrium one can define the normalized amplitudes $A_I:=\phi_I(\mathbf{r}) A_I^0$, where $\phi_I(\mathbf{r}) \in [0,1]$ and $A_I^0$ denotes the equilibrium amplitudes: $A_1^0=a_1 \sqrt{\epsilon}$ and $A_{I>1}^0=a_I \epsilon^{3/2}$. Note that for a planar equilibrium interface $\phi_I(x\to\pm\infty)\to 0,1$, respectively. With this re-scaling, the equilibrium bulk liquid and solid phases are described by $\vec{\phi}_L=(0,0,0,\dots)$ and $\vec{\phi}_S=(1,1,1,\dots)$, respectively. Considering only the leading order terms of Eq. (\ref{eq:expand}) and substituting $A_1=A_1^0 \phi$ and $\bar{\psi}=c_\psi \sqrt{\epsilon}$ yields
\begin{equation}
\label{eq:genGL}
\frac{\Delta f(\phi)}{\epsilon^2} = \left( \frac{3 c_\psi^2-1}{2} N_1 a_1^2 \right) \phi^2 + c_\psi N_3 a_1^3 \phi^3 + \frac{N_4}{4}a_1^4 \phi^4 \enskip,
\end{equation}
where $c_\psi$ and $a_1$ are defined by  Eq. (\ref{eq:coefficient1}) and (\ref{eq:coefficient2}). Substituting these into Eq. (\ref{eq:genGL}) yields
\begin{equation}
\label{eq:funcGL}
\frac{\Delta f(\phi)}{\epsilon^2} = \tilde{w} [\phi(1-\phi)]^2 \enskip ,
\end{equation}
where
\begin{equation}
\tilde{w}=\frac{4 N_1^2 N_3^4}{N_4 (3N_1N_4-2N_3^2)^2} \enskip .
\end{equation}
Note that Eq. (\ref{eq:funcGL}) is exactly the well-known $4^{th}$ order Ginzburg-Landau polynomial for triangular and bcc structures. (For the fcc structure $N_3\equiv 0$, therefore, there is no fcc-liquid first-order phase transition in the Swift-Hohenberg formalism in leading order, i.e. close to the critical point.) 

\subsubsection{Amplitude equation}

The isotropic single order parameter amplitude equation in equilibrium can be written as:
\begin{equation}
\label{eq:amplitude_iso}
F_{\rm iso} = \int dV \left\{ \kappa (\nabla\phi)^2 + w g(\phi) \right\} \enskip ,
\end{equation}
where $g(\phi)=[\phi(1-\phi)]^2$ is defined by Eq. (\ref{eq:funcGL}) and $w=\tilde{w}\epsilon^2$. The equilibrium solution of the Euler-Lagrange equation $\delta F_{\rm iso}/\delta \phi=0$ is the kink-function $\phi(x)=[1-\tanh(x/d)]/2$, where $d=2\sqrt{\kappa/w}$. The interfacial free energy can be obtained by using the integral Euler-Lagrange equation: $\gamma = \sqrt{\kappa w}/3$. The model parameters $\kappa$ and $w$ can be then related to the interfacial free energy and interface thickness as:
\begin{equation}
w = 6(\gamma/d) \quad \text{and} \quad \kappa=(3/2)\gamma d \enskip .
\end{equation}
Substituting Eqns. (\ref{eq:isogamma}) and (\ref{eq:isolambda}) into the above equation yields
\begin{eqnarray}
\label{eq:amplitudew}
\frac{w}{\epsilon^2} &=& 4 \left(\frac{N_1^2}{N_4}\right) \left(\frac{N_3^2}{3N_1N_4-2N_3^2}\right)^2 \\
\label{eq:amplitudekappa}
\frac{\kappa}{\epsilon} &=& 8C \left(\frac{N_1^2}{N_4}\right) \left( \frac{N_3^2}{3N_1N_4-2N_3^2}  \right) \enskip .
\end{eqnarray} 
Note that Eq. (\ref{eq:amplitudew}) consistently recovers Eq. (\ref{eq:funcGL}), showing that our calculation is self-consistent. Also note that Eq. (\ref{eq:amplitude_iso}) consistently verifies the divergence of the interface thickness found in Eq. (24).

\subsection{Anisotropic extension}

Introducing the order parameters $\phi_i(\mathbf{r}):=A_i^1/A_1^0 \in [0,1]$ gives the anisotropic extension of Eq. (\ref{eq:genGL}):
\begin{equation}
\label{eq:anisoGL}
\begin{split}
g(\vec{\phi}):= & \frac{\Delta f(\{\phi_i\})}{w} = \frac{1}{N_1}\sum_{i} \phi_i^2 - \\
& - \frac{2}{N_3} \sum_{i,j,k}  (\phi_i \phi_j \phi_k) \delta^{1,1,1}_{i,j,k} + \\
& + \frac{1}{N_4} \sum_{i,j,k,l}  (\phi_i \phi_j \phi_k \phi_l) \delta^{1,1,1,1}_{i,j,k,l} \enskip ,
\end{split}
\end{equation}
Following Majaniemi and Provatas \cite{PhysRevE.79.011607}, for a planar interface the anisotropic interfacial free energy can be written as:
\begin{equation}
\label{eq:gamma_aniso_gl}
\gamma(\mathbf{n}) = \int_{-\infty}^{\infty} d\xi \left\{ \kappa \left[ (\partial_\xi \vec{\phi})^T \cdot \mathbb{S}(\mathbf{n}) \cdot (\partial_\xi\vec{\phi}) \right] + \Delta f [\vec{\phi}(\xi)] \right\} \enskip ,
\end{equation}
where $ \Delta f [\vec{\phi}] $  is defined by Eq. (\ref{eq:anisoGL}). Here $\kappa$ and $w$ are defined by Eqns. (\ref{eq:amplitudew}) and (\ref{eq:amplitudekappa}) again. The elements of the coefficient matrix $\mathbb{S}(\mathbf{n})$ can be determined by substituting $\phi^*_i(\xi) = [1+g_i(\xi)]/2$ into Eq. (\ref{eq:gamma_aniso_gl}), and comparing the result with Eq. (\ref{eq:gamma_simp}) after substituting $A_1=a_1\sqrt{\epsilon}$, $\bar{\psi}=c_\psi\sqrt{\epsilon}$ with Eqns. (\ref{eq:coefficient1}) and (\ref{eq:coefficient2}), and considering Eqns. (\ref{eq:amplitudew}) and (\ref{eq:amplitudekappa}) (for details, see Appendix E). The calculation then yields a diagonal matrix $s_{ij}=\delta_{ij}s_i$ with elements $s_i = \zeta(\mathbf{n}\cdot\mathbf{\Gamma}_i^1)/(N_1 C)$. Finally, the corresponding anisotropic Ginzburg-Landau free energy functional then reads as:
\begin{equation}
\label{eq:amplitude_aniso}
F_{\rm aniso} = \int dV \left\{ \kappa \sum_i (\nabla \phi_i)^T \cdot \mathbb{A}_i \cdot (\nabla \phi_i) + w \cdot g(\vec{\phi}) \right\} \enskip ,
\end{equation}
where 
\begin{equation}
\mathbb{A}_i = \frac{1}{N_1 C}\left [ \zeta_0 \cdot \mathbb{I} + \zeta_1 \cdot (\mathbf{\Gamma}_i^1 \otimes \mathbf{\Gamma}_i^1) \right] \enskip ,
\end{equation}
 while $g(\vec{\phi})$, $\kappa$ and $w$ are defined by Eqns. (\ref{eq:anisoGL}), (\ref{eq:amplitudew}) and (\ref{eq:amplitudekappa}), respectively. Note that $\kappa$ and $w$ scale out from the free energy functional [and also from Eq. (\ref{eq:amplitude_aniso_gamma_base})], therefore, the anisotropy of the interfacial free energy is constant and depends exclusively on the crystal structure [the form of $g(\vec{\phi})$], while its magnitude is just the magnitude of the critical point remnant anisotropy of the Phase-Field Crystal model [Eq. (\ref{eq:amplitude_aniso_gamma_base}) shows the leading order of the interfacial free energy close to the critical point].

\section{Conclusions}

Our calculations show that a weakly $4^{th}$-order anisotropic one-mode Ginzburg-Landau theory inherits the properties of a leading-order amplitude model of a one-mode dominant $4^{th}$-order classical Density Functional Theory close to its critical point. The constant anisotropy appearing in weakly $4^{th}$ order Ginzburg-Landau theories originates from the fact that all material parameters (except the crystal structure) scale out from the free energy functional in the determination of the crystalline anisotropy. We have shown that the magnitude of the GL anisotropy coincides with the critical point remnant anisotropy of the generating density functional theory. We have to emphasize that the non-vanishing behavior of the anisotropy doesn't contradict to the mean-field theory, since the anisotropy is a secondary quantity, i.e. its critical exponent is not related to the important exponents.

Our results have consequences on the quantitative applicability of both the Phase-Field Crystal model and Ginzburg-Landau theories emerging from it. In the case of the PFC model the numerical calculations resulted in a remnant ($\nu_0 \approx 3\%$) anisotropy in the range $0 < \epsilon \lesssim 0.1$. In this range $d_{10\%-90\%} \gtrsim 3\sigma_0$, where $d_{10\%-90\%}$ is the usual $10\%-90\%$ interface thickness and $\sigma_0$ the bcc lattice constant. Since this is true for simple metals, \textit{$\epsilon$ is not a relevant parameter in quantifying the anisotropy for metallic materials}. In contrast, it has been found that $\nu_0$ inherited by the GL theory exclusively depends on the form of the scaled direct correlation function $\hat{c}_2$. Since the symmetry breaking of the GL coefficient matrix is trivially related to properties of the direct correlation function, one can calibrate the anisotropy in the Ginzburg-Landau theory by investigating the critical behavior of the generating PFC.\\

A possible pathway of deriving consistent GL theories, in accordance with the original idea of Shih et al. \cite{PhysRevA.35.2611}, is to choose such a PFC description, in which more than one RLV set is dominant, i.e. we at least two peaks of the direct correlation function are considered. The best candidate is the so-called structural PFC (or XPFC) model \cite{PhysRevLett.105.045702}, in which the peak peak heights are weighted by the Debye-Waller factor. Since the peak heights are not equal, the critical point vanishes, meaning that the $\epsilon$ dependence appears in the amplitude theory. Nevertheless, combining the XPFC model with the recently published fluctuating hydrodynamic theory of freezing \cite{0953-8984-26-5-055001} might result in a continuum description of crystallization of simple liquids on the (classical) fundamental length scale of the material. Moreover, comparing the results of the model with molecular dynamics data will hopefully anchor $\epsilon$ to the physical temperature, making the model fully quantitative.

\section*{Acknowledgement} 
The authors wish to thank professor L. Gr\'an\'asy and F. Podmaniczky from the Wigner Research Centre for Physics, Hungary, for the valuable discussions which significantly contributed to the quality of the work. This work has been supported by the Postdoctoral Programme of the Hungarian Academy of Sciences and the Natural Sciences and Engineering Research Council of Canada.

\bibliography{./references}

\section*{Appendix A: Evaluation of the bulk free energy density}

In order to evaluate Eq. (\ref{eq:dfdensdef}) for $\psi_p(\mathbf{r})=\bar{\psi}+\Delta \psi(\mathbf{r})$, where $\Delta \psi(\mathbf{r})=\sum_I A_I \sum_{i \in S(I)} \exp^{\imath \mathbf{\Gamma}_i^I \cdot \mathbf{r}}$, first we re-formulate Eq. (\ref{eq:model_base}) as follows:
\begin{equation}
\label{eq:appa_base}
\mathcal{F} = \int d\mathbf{r} \left\{ \frac{1}{2} \sum_{n=0}^{\infty} \alpha_n (\nabla^n \psi)^2 -\epsilon \frac{\psi^2}{2} + \frac{\psi^4}{4} \right\} \enskip ,
\end{equation}
where we used that the functional derivative
\begin{equation}
\frac{\delta \mathcal{F}}{\delta \psi} = \sum_{i=0}^\infty (-1)^i \frac{\partial I}{\partial \nabla^i \psi}
\end{equation}
results in the same for both $\psi \cdot \hat{c}_2[ \psi]=\sum_{n=0}^{\infty} \alpha_n \psi [(-\nabla^2)^n \psi] $ and $\sum_{n=0}^{\infty} \alpha_n (\nabla^n \psi)^2$. The spatial derivatives of $\psi(\mathbf{r})$ read as:
\begin{equation}
\nabla^n \psi(\mathbf{r}) =  \sum_I A_I \sum_{i \in S(I)} (\imath  \mathbf{\Gamma}_i^I )^n \exp^{\imath \mathbf{\Gamma}_i^I \cdot \mathbf{r}} \enskip ,
\end{equation}
where $n>1$. Introducing the shorthand notation $\langle . \rangle := \frac{1}{V_{cell}} \int_{V_{cell}} dV \{ . \}$ for the lattice cell average the following terms emerge from $\psi \cdot \hat{c}_2[\psi]$ in the free energy density:
\begin{equation}
\langle (\nabla^n \psi)^2 \rangle = \sum_{I,J} A_I A_J \sum_{i,j} [- \mathbf{\Gamma}_i^I \cdot \mathbf{\Gamma}_j^J]^n \left\langle \exp^{\imath (\mathbf{\Gamma}_i^I+\mathbf{\Gamma}_j^J) \cdot \mathbf{r}} \right\rangle \enskip ,
\end{equation}
where
\begin{equation}
\left\langle \exp^{\imath (\mathbf{\Gamma}_i^I+\mathbf{\Gamma}_j^J) \cdot \mathbf{r}} \right\rangle = \delta(\mathbf{\Gamma}_i^I+\mathbf{\Gamma}_j^J)
\end{equation} 
is the (Kronecker) delta-function giving $1$ for $\mathbf{\Gamma}_i^I=-\mathbf{\Gamma}_j^J$, and $0$ otherwise. Therefore,
\begin{equation}
\langle (\nabla^n \psi)^2 \rangle = \sum_I A_I^2 \mathcal{N}^{(2)}_{I,I} (\Gamma_I)^{2n} \enskip ,
\end{equation} 
where $\mathcal{N}^{(2)}_{I,I}=\sum_{i,j} \delta(\mathbf{\Gamma}_i^I+\mathbf{\Gamma}_j^J)$ is just the number of RLVs in the $I^{th}$ RLV set. Furthermore,
\begin{equation}
\langle \psi^2 \rangle =  \bar{\psi}^2+ \langle \Delta\psi^2 \rangle = \bar{\psi}^2+ \sum_I A_I^2 \mathcal{N}_{I,I}^{(2)} \enskip ,
\end{equation}
where we used that $\langle \Delta \psi \rangle=0$. 
Finally,
\begin{equation}
\frac{1}{2}\sum_{n=0}^\infty \alpha_n \langle (\nabla^n \psi)^2 \rangle = \alpha_0 \frac{\bar{\psi}^2}{2} + \frac{1}{2} \sum_I A_I^2 \mathcal{N}^{(2)}_{I,I} \sum_{n=0}^\infty \alpha_n (\Gamma_I)^{2n} \enskip .
\end{equation}
Note that $\sum_{n=0}^\infty \alpha_n (\Gamma_I)^{2n} \equiv c_2(\Gamma_i)$. Then, the contribution of $\psi \cdot \hat{c}_2[\psi]$ to the free energy density reads as:
\begin{equation}
\frac{1}{2} \langle \psi \cdot \hat{c}_2[\psi]\rangle = \alpha_0 \frac{\bar{\psi}^2}{2} + \frac{1}{2} \sum_I A_I^2 \mathcal{N}^{(2)}_{I,I} c_2(\Gamma_I) \enskip .
\end{equation}
Introducing $\mathcal{N}_{I,J,K}^{(3)}:=\sum_{i,j,k}\delta(\mathbf{\Gamma}_i^I+\mathbf{\Gamma}_j^J+\mathbf{\Gamma}_k^K)$ and $\mathcal{N}_{I,J,K,L}^{(4)}:=\sum_{i,j,k,l}\delta(\mathbf{\Gamma}_i^I+\mathbf{\Gamma}_j^J+\mathbf{\Gamma}_k^K+\mathbf{\Gamma}_l^L)$, where $i \in S(I), j \in S(J), k \in S(K)$ and $l \in S(L)$, and taking into account that
\begin{equation}
\langle \psi^4 \rangle = \bar{\psi}^4 + 4 \bar{\psi} \langle \Delta\psi^3 \rangle + 6 \bar{\psi}^2 \langle \Delta\psi^2 \rangle + \langle \Delta\psi^4 \rangle \enskip ,
\end{equation}
where
\begin{equation}
\langle \Delta \psi^3 \rangle = \sum_{I,J,K} A_I A_J A_K \mathcal{N}_{I,J,K}^{(3)} \enskip ,
\end{equation}
and
\begin{equation}
\langle \Delta \psi^4 \rangle = \sum_{I,J,K,L} A_I A_J A_K A_L \mathcal{N}_{I,J,K,L}^{(4)}
\end{equation}
yields
\begin{equation}
\begin{split}
f[\psi_p] = &\sum_{I} \left[ A_I^2 N^{(2)}_{I,I} \right] \frac{c_2(\Gamma_I) - \epsilon + 3\bar{\psi}^2}{2} + \\
& + \bar{\psi} \sum_{I,J,K} (A_I A_J A_K) \mathcal{N}^{(3)}_{I,J,K} \\
& + \frac{1}{4}\sum_{I,J,K,L} (A_I A_J A_K A_L) \mathcal{N}^{(4)}_{I,J,K,L} + f[\bar{\psi}] \enskip ,
\end{split}
\end{equation}
where $f[\bar{\psi}]=(\alpha_0-\epsilon)(\bar{\psi}^2/2)+\bar{\psi}^4/4$. Therefore, $\Delta f[\psi_p]=f[\psi_p]-f[\bar{\psi}]$ results in Eq. (\ref{eq:dfbase}).

\section*{Appendix B: INCLUDING THE EQUILIBRIUM DENSITY JUMP}

If one includes the equilibrium crystal-liquid density jump, the relevant thermodynamic potential is grand potential density, which reads as:
\begin{eqnarray}
\omega_s &:=& f[\psi_p] - \mu_s \psi_s \\
\omega_l &:=&  f[\psi_l] - \mu_l \psi_l
\end{eqnarray}
where $\psi_p(\mathbf{r})=\psi_s+\Delta\psi(\mathbf{r})$ is the bulk solid solution, where $\Delta \psi(\mathbf{r})=\sum_I A_I \sum_{i \in S(I)} \exp(-\imath \mathbf{r}\cdot \mathbf{\Gamma}_i^I)$, while $\psi_s$ and $\psi_l$ are the equilibrium average densities of the crystal and the liquid, respectively. The chemical potential reads as
\begin{equation}
\mu(\psi) = \left.\frac{\delta F}{\delta \psi}\right|_\psi \enskip .
\end{equation}
 In this case, the equilibrium condition comes from the common tangent construction:
 \begin{equation}
 \label{eq:equi_consider}
\Delta \omega = \omega_s - \omega_l = 0 \enskip , \quad \text{and} \quad \mu_s=\mu_l \enskip .
 \end{equation}
These two equations define the equilibrium solid and liquid densities, $\psi_s$ and $\psi_l$, respectively. Considering the $0^{th}$-order of $\Delta \omega=0$ and $\partial f_s/\partial A_I=0$ yields $q_0=1$, $y_s=y_1=1/2$ and $y_I=3/2$ for any $I>1$, where $y_s$ is the critical exponent of the solid equilibrium density, i.e. $\psi_s=c_s \cdot \epsilon^{y_s}$. Using these, $\mu_s=\mu_l$ starts with
\begin{eqnarray}
(1-\epsilon)(\psi_s-\psi_l)=0 \quad \Rightarrow \quad y_l=1/2, \enskip c_l=c_s \enskip ,
\end{eqnarray}
where $\psi_l=c_l \cdot \epsilon^{y_l}$ is the equilibrium liquid density. Note that $\psi_l$ and $\psi_s$ are equal in the leading order, i.e. $\psi_{s,l}=c_\psi \cdot \epsilon^{y_\psi}$ again, where $y_\psi=1/2$. Therefore, $y_\Delta>1/2$ in $\Delta := \psi_s - \psi_l = 2 c_\Delta \cdot \epsilon^{y_\Delta}$. Using $\psi_l=\bar{\psi}-\delta$ and $\psi_s=\bar{\psi}+\delta$ [where $\bar{\psi}:=(\psi_l+\psi_s)/2=c_\psi \cdot \epsilon^{y_\psi}$ and $\delta:=(\psi_s-\psi_l)/2=c_\Delta \cdot \epsilon^{y_\Delta}$] in the next order of the equilibrium condition $\Delta \omega=0$ yields
\begin{equation}
y_\Delta = 3/2 \enskip , 
\end{equation}
which is the known mean-field result for crystal-liquid phase transitions. 

\section*{Appendix C: LEADING ORDER OF THE ANISOTROPIC INTERFACIAL FREE ENERGY}

\subsection*{I. Neglecting the equilibrium density jump}

In order to evaluate the interfacial free energy, first we modify Eq. (\ref{eq:gamma_base}) as follows:
\begin{equation}
\label{eq:tension_definition}
\gamma(\mathbf{n}) :=  \int_{-\infty}^{\infty} d\xi \left( \frac{1}{A_\perp}\int_\xi dA_\perp \left\{ \Delta I[\psi_{sl}] - \tau \cdot \Delta I[\psi_p] \right\} \right),
\end{equation}
where $\tau$ is to be determined later. Note that this modification is purely formal, since the contribution from $\Delta I[\psi_p]$ vanishes because of the equilibrium condition:
\begin{equation*}
\begin{split}
&\int_{-\infty}^{\infty} d\xi \left( \frac{1}{A_\perp}\int_\xi dA_\perp \left\{ \Delta I[\psi_{p}(\mathbf{r})] \right\} \right) \propto \\
& \propto \int dV \{\Delta I[\psi_p]\} \propto \Delta f[\psi_p] \equiv 0 \enskip .
\end{split}
\end{equation*}
For the sake of simplicity, first we introduce the shorthand notation
\begin{equation*}
\left\langle . \right\rangle_\xi:=(1/A_\perp) \int_\xi dA_\perp \{.\} \enskip .
\end{equation*}
Substituting Eq. (\ref{eq:slprof}) into Eq. (\ref{eq:gamma_base}), and using Eq. (\ref{eq:subsgh}), the terms appearing in the interfacial free energy can be expressed in the following general form:
\begin{equation}
\label{eq:gen_term}
\int_{-\infty}^{\infty} d\xi \left\langle \prod_{I,i} \partial^{D_i^I} g_i^I(\xi) \prod_{J,j} h_j^J(\mathbf{r}) \right\rangle_\xi \enskip ,
\end{equation}
where $\partial^{D_i^I} g_i^I(\xi) = [\partial^{D_i^I}/\partial (\xi/\Lambda_i^I)^{D_i^I}][\tanh(\xi/\Lambda_i^I)]$, and $\prod_{I,i}$ runs for some arbitrary RLVs. In order to evaluate Eq. (\ref{eq:gen_term}) first we decompose the coordinate as $\mathbf{r}=\xi \cdot \mathbf{n}+\mathbf{r}_\perp$, where $\mathbf{r}_\perp \cdot \mathbf{n} \equiv 0$ (in other words, $\mathbf{r}_\perp$ is in the interface). Using this in Eq. (\ref{eq:subsgh}) results in:
\begin{equation}
\label{eq:gen_term_decomp}
\mathcal{I}_\perp \cdot \int_{-\infty}^{\infty} d\xi \left\{  \prod_{I,i} \partial^{D_i^I} g_i^I(\xi)  \prod_{J,j} h_j^J(\xi \cdot \mathbf{n}) \right\} \enskip ,
\end{equation}
where
\begin{equation}
\label{eq:kintegral}
\mathcal{I}_\perp = \left\langle \exp \left[ \imath \left( \sum_{J,j} \mathbf{\Gamma}_j^J \right)\mathbf{r}_\perp \right] \right\rangle_\xi = \delta_{j_1,j_2,\dots,j_N}^{J_1,J_2,\dots,J_N}
\end{equation}
is \textit{not} a function of $\xi$. Here we used the shorthand notation $\delta_{j_1,j_2,\dots,j_N}^{J_1,J_2,\dots,J_N}:=\delta \left( \sum_{n=1}^{N} \mathbf{\Gamma}_{j_n}^{J_n} \right)$. Note that if Eq. (\ref{eq:kintegral}) gives 1, then $\prod_{J,j} h_j^J(\xi \cdot \mathbf{n})$ also gives 1, otherwise Eq. (\ref{eq:gen_term_decomp}) is equal to 0. Therefore, the term $\prod_{J,j} h_j^J(\xi \cdot \mathbf{n})$ in Eq. (\ref{eq:gen_term_decomp}) can be omitted. Using the shorthand notation $\| . \| :=\int_{-\infty}^{\infty}d\xi\{.\}$ Eq. (\ref{eq:gen_term}) can be re-written as:
\begin{equation}
\label{eq:integral_rule}
\begin{split}
&\int_{-\infty}^{\infty} d\xi \left\langle \prod_{I,i} \partial^{D_i^I} g_i^I(\xi) \prod_{J,j} h_j^J(\mathbf{r}) \right\rangle_\xi  = \\
& = \| \prod _{I,i}\partial^{D_i^I} g_i^I(\xi) \| \cdot \delta_{j_1,j_2,\dots,j_N}^{J_1,J_2,\dots,j_N} \enskip .
\end{split}
\end{equation}
Note that this derivation is true only if $\mathbf{k} := \sum_j \mathbf{\Gamma}_j^J \neq 0$ is not parallel with $\mathbf{n}$, otherwise $\mathbf{r}_\perp \cdot \mathbf{k} \equiv 0$. In this case correction terms emerge, however, it can be shown that they vanish for $\epsilon \to 0$ (The proof is beyond the scope of this paper.).

Following the derivation presented in Appendix A, we can evaluate Eq. (\ref{eq:tension_definition}) as follows: First we introduce the shorthand notation $\langle . \rangle_{\gamma} := \| \langle . \rangle_\xi \|$. Considering $\Delta I[\psi]=I[\psi]-I[\bar{\psi}]$, where $I[.]$ is the integrand of Eq. (\ref{eq:appa_base}) yields
\begin{equation}
\begin{split}
& \langle \Delta I[\psi] \rangle_\gamma = \frac{1}{2}\sum_{n=0}^\infty \alpha_n \langle (\nabla^n \Delta \psi)^2 \rangle_\gamma + \\
& + \frac{3 \bar{\psi}^2-\epsilon}{2}\langle \Delta\psi^2 \rangle_\gamma + \bar{\psi} \langle \Delta\psi^3 \rangle_{\gamma} + \frac{1}{4} \langle \Delta\psi ^4 \rangle_{\gamma} \enskip ,
\end{split}
\end{equation}
where $\psi$ can be either $\psi_{sl}$ or $\psi_p$. Introducing $\psi_{sl}=\bar{\psi}+\Delta\psi_{sl}$ and $\psi_p=\bar{\psi}+\Delta\psi_p$ in Eq. (\ref{eq:tension_definition}) results in
\begin{equation}
\label{eq:appa_expand}
\begin{split}
\gamma(\mathbf{n}) &=  \langle \Delta I[\psi_{sl}] \rangle_\gamma -\tau \langle \Delta I[\psi_p] \rangle_\gamma = \\
&=  \frac{1}{2}\sum_{n=0}^\infty \alpha_n \langle (\nabla^n \Delta \psi_{sl})^2 \rangle_\gamma - \tau \langle (\nabla^n \Delta \psi_p)^2 \rangle_\gamma + \\
& + \frac{3 \bar{\psi}^2-\epsilon}{2} \left( \langle \Delta\psi_{sl}^2 \rangle_\gamma- \tau \langle \Delta\psi_p^2 \rangle_\gamma \right) + \\
& + \bar{\psi} \left( \langle \Delta\psi_{sl}^3 \rangle_\gamma - \tau \langle \Delta\psi_p^3 \rangle_{\gamma} \right) + \\
& + \frac{1}{4} \left( \langle \Delta\psi_{sl}^4 \rangle_\gamma- \tau \langle \Delta\psi_p^4 \rangle_{\gamma} \right) \enskip .
\end{split}
\end{equation}
Using $\Delta\psi_{sl}=\sum_I A_I \sum_{i \in S(I)} [(1+g_i^I)/2]h_i^I$, $\Delta\psi_p=\sum_I A_I \sum_{i \in S(I)} h_i^I$ together with Eq. (\ref{eq:integral_rule}) in Eq. (\ref{eq:appa_expand}), and choosing $\tau=1/2$ yields
\begin{eqnarray*}
\langle&& \Delta\psi_{sl}^2 \rangle_\gamma- \frac{1}{2} \langle \Delta\psi_p^2 \rangle_\gamma = \frac{1}{4}\sum_I A_I^2 \sum_{i \in S(I)} \| (g_i^I)^2-1\| \\
 \langle&& \Delta\psi_{sl}^3 \rangle_\gamma- \frac{1}{2} \langle \Delta\psi_p^3 \rangle_{\gamma} = \\
&& = \frac{1}{8}\sum_{I,J,K} A_I A_J A_K \sum_{i,j,k} \sum_{(m,n)}^{(i,j,k)}\| g_m^M g_n^N -1\| \delta_{i,j,k}^{I,J,K} \\
\langle&& \Delta\psi_{sl}^4 \rangle_\gamma- \frac{1}{2} \langle \Delta\psi_p^4 \rangle_{\gamma} = \\ 
&& = \frac{1}{16}\sum_{I,J,K} A_I A_J A_K A_L \left( \sum_{i,j,k,l} \|g_i^I g_j^J g_k^K g_l^L -1 \| + \right . \\
&& \left. + \sum_{(m,n)}^{(i,j,k,l)}\| g_m^M g_n^N -1\| \right ) \delta_{i,j,k,l}^{I,J,K,L} \enskip .\\
\end{eqnarray*}
To find the first term of Eq. (\ref{eq:appa_expand}) we write
\begin{eqnarray}
\label{eq:fckbig}
\nonumber \langle && (\nabla \Delta\psi_{sl})^2 - (\nabla \Delta\psi_{p})^2 \rangle_\gamma = \\
\nonumber &&= \frac{1}{4} \sum_{I} A_I^2 \sum_{i \in S(I)} \left[ \|(g_i^I)^2-1\|\Gamma_I^2 + \frac{\|(\partial g_i^I)^2\|}{(\Lambda_i^I)^2}\right] \\
\nonumber \langle && (\nabla^2 \Delta\psi_{sl})^2 - (\nabla^2 \Delta\psi_{p})^2 \rangle_\gamma = \\
\nonumber &&= \frac{1}{4} \sum_{I} A_I^2 \sum_{i \in S(I)} \left[ \|(g_i^I)^2-1\|\Gamma_I^4 - 2\frac{\|g_i^I \partial^2 g_i^I\|}{(\Lambda_i^I)^2}\Gamma_I^2 +\right. \\
\nonumber && \left. + 4(\mathbf{n}\cdot \mathbf{\Gamma}_i^I)^2 \frac{\|(\partial g_i^I)^2\|}{(\Lambda_i^I)^2} + O(1/\Lambda^3) \right] \\
\nonumber \langle && (\nabla^3 \Delta\psi_{sl})^2 - (\nabla^3 \Delta\psi_{p})^2 \rangle_\gamma = \\
\nonumber &&= \frac{1}{4} \sum_{I} A_I^2 \sum_{i \in S(I)} \left[ \|(g_i^I)^2-1\|\Gamma_I^6 + \frac{\| (\partial g_i^I)^2 \|}{(\Lambda_i^I)^2}\Gamma_I^4 +\right. \\
\nonumber && \left. + 8(\mathbf{n}\cdot \mathbf{\Gamma}_i^I)^2 \Gamma_I^2 \frac{\|(\partial g_i^I)^2\|}{(\Lambda_i^I)^2} + O(1/\Lambda^3) \right] \\
\nonumber \langle && (\nabla^n \Delta\psi_{sl})^2 - (\nabla^n \Delta\psi_{p})^2 \rangle_\gamma = \\
\nonumber &&= \frac{1}{4} \sum_{I} A_I^2 \sum_{i \in S(I)} \left[ \|(g_i^I)^2-1\|\Gamma_I^{2n} + \right . \\
&& \left. + \left( \propto \frac{\| . \|}{(\Lambda_i^I)^2} \right) + O(1/\Lambda^3) \right] \enskip .
\end{eqnarray}
Using that $\|(g_i^I)^2-1\|=-2\Lambda_i^I$, $\|(\partial g_i^I)^2\|=(4/3)\Lambda_i^I$, or $\|.\| \propto \Lambda_i^I$ in general (this is trivial since all the functions have the same argument, i.e. $\xi/\Lambda_i^I$), and substituting all the terms into Eq. (\ref{eq:appa_expand}) yields:
\begin{equation}
\label{eq:appb_gamma}
\begin{split}
\gamma(\mathbf{n}) = & \sum_{I} A_I^2 \sum_{i \in S(I)} \left\{ \frac{2 \zeta(\mathbf{n}\cdot\mathbf{\Gamma}_i^I,\Gamma_I)}{3 \Lambda_i^I} - \right. \\
& \left. - \frac{c_2(\Gamma_I)-\epsilon+3\bar{\psi}^2}{4} \Lambda_i^I  \right\} +\\
& \frac{\bar{\psi}}{8} \sum_{I,J,K} A_I A_J A_K \sum_{i,j,k} \left[ \sum_{(m,n)}^{(i,j,k)} \|g_m^M g_n^N -1 \| \right] \delta_{i,j,k}^{I,J,K} + \\
& \frac{1}{64} \sum_{I,J,K,L} A_I A_J A_K A_L \sum_{i,j,k,l} \left[ \| g_i^I g_j^J g_k^K g_l^L -1 \| + \frac{}{} \right. \\
& \left. + \sum_{(m,n)}^{(i,j,k,l)} \| g_m^M g_n^N -1\| \right] \delta_{i,j,k,l}^{I,J,K,L} \enskip ,
\end{split} 
\end{equation}
where we neglected the terms in the order of $1/(\Lambda_i^I)^3]$. Here $\zeta(\mathbf{n}\cdot\mathbf{\Gamma}_i^I,\Gamma_I)$ collects all terms proportional to $1/\Lambda_i^I$ of Eq. (\ref{eq:fckbig}). Note that the $c_2(\Gamma_I)$ term comes from the sum $\sum_{n=0}^\infty \alpha_n \|(g_i^I)^2-1\| \Gamma_I^{2n}=(-2\Lambda_i^I)c_2(\Gamma_I)$. Taking into account the result of Appendix D, i.e. that the critical exponents of the characteristic interface thicknesses must be equal, the leading order term of Eq. (\ref{eq:appb_gamma}) is precisely Eq. (\ref{eq:gamma_simp}).

\subsection*{II. Including the equilibrium density jump}

Repeating the calculation for the case when the crystal-liquid equilibrium density jump is also considered is straightforward. In this case we use the definition of the surface tension:
\begin{equation}
\label{eq:tension_definition}
\begin{split}
\gamma'(\mathbf{n}) := &  \int_{-\infty}^{\infty} d\xi \left( \frac{1}{A_\perp}\int_\xi dA_\perp \left\{ \Delta I'[\psi_{sl}] - \right. \right .\\
& \left. \left. - \frac{1}{2} \cdot (\Delta I'[\psi_s] + \Delta I'[\psi_l] \right\} \right).
\end{split}
\end{equation}
Here $\Delta I'[\psi]=I'[\psi] - I'[\psi_l]$, where $I'[\psi]=I[\psi]-\mu \cdot \psi$ and $\mu=\delta F/\delta \psi$. Note that $\Delta I'[\psi_l]\equiv 0$ and $(1/A_\perp)\int dV \Delta I'[\psi_s] \propto \Delta\omega[\psi_s]=0$ is the equilibrium condition. Furthermore, we use the following approximations:
\begin{eqnarray*}
\psi_{sl} &=& \bar{\psi} + \Delta\varphi \sum_{i \in S(1)} g_i^1 + \sum_{i \in S(1)} \frac{1+g_i^1}{2} h_i^1 \\
\psi_{s} &=& \bar{\psi} + \frac{\Delta}{2} + \sum_{i \in S(1)} h_i^1 \enskip , \\
\psi_{l} &=& \bar{\psi} - \frac{\Delta}{2} \\
\end{eqnarray*}
where $\Delta\varphi=\Delta/N_1$, i.e. the density jump $\Delta$ is distributed equally between the $N_1$ RLV vectors of the dominant RLV set. After a lengthy but straightforward calculation one can conclude to
$\gamma'(\mathbf{n}) =  \gamma(\mathbf{n}) + O(\Delta)$, where $\gamma(\mathbf{n})$ is defined by Eq. (\ref{eq:appb_gamma}) and $O(\Delta) \propto \epsilon^{5/2}$. Therefore, the equilibrium density jump has no contribution to the leading order of the interfacial free energy.

\section*{Appendix D: APPROXIMATING THE ENVELOPE FUNCTION INTEGRALS}

In order to investigate the general, anisotropic case first we have to calculate $\| g_i^I g_j^J -1\|$ and $\| g_i^I g_j^J g_k^K g_l^L -1\|$ in Eq. (\ref{eq:gamma_simp}). Unfortunately, \textit{no analytical formulae are known} for these integrals as a function of the parameters $\Lambda_i^1,\Lambda_j^1,\Lambda_k^1$ and $\Lambda_l^1$. However, we can start  from the integral $\|g_i^I g_j^j-1\|$:
\begin{equation}
\begin{split}
&\Lambda_i^I \int_{-\infty}^\infty dx \left[ \tanh(x)\tanh\left( \frac{\Lambda_i^I}{\Lambda_j^J} x \right)-1 \right] \equiv \\
& \equiv \Lambda_j^J \int_{-\infty}^\infty dy \left[ \tanh\left( \frac{\Lambda_j^J}{\Lambda_i^I} y \right)\tanh(y)-1 \right] \enskip .
\end{split}
\end{equation} 
Introducing $f(\eta):=\int dx [\tanh(x)\tanh(\eta x)-1]$, where $\eta=\Lambda_i^I/\Lambda_j^J$ yields the following general constraint:
\begin{equation}
\label{eq:define_integral}
\eta \cdot f[\eta] = f(1/\eta) \enskip , \quad \text{and} \quad f(1)=-2 \enskip , 
\end{equation}
which defines a family of functions for $f(\eta)$. More generally, we can use the following Ansatz:
\begin{equation}
\label{eq:tanh_fit}
\| \prod_{k}^n g_{i_k}^{I_k} -1 \| \approx \sum_{s}^m \frac{f^{(n)}_s}{\mathcal{N}[\mathbf{p}_s]} \sum_{l \in \mathcal{P}[\mathbf{p}_s]} \prod_{k}^n \left( \Lambda_{i_k}^{I_k} \right)^{p_{k(l)}^{(s)}} \\
\end{equation}
where $\sum_{l \in \mathcal{P}[\mathbf{p}_s]}$ runs over all permutations of the power set $\mathbf{p}_s=\left\{ p_1^{(s)},p_2^{(s)},\dots,p_n^{(s)} \right\}$ [i.e. $p_{k(l)}^{s}$ denotes the $k^{th}$ element in the $l^{th}$ permutation of $\mathbf{p}_s$], $\sum_{k=1}^n p_k^{(s)}=1$, $\mathcal{N}[\mathbf{p}_s]$ is the number of permutations, and the fitting parameters satisfy
\begin{equation}
\sum_s f_s^{(n)} = \int_{-\infty}^{\infty} dx \{ \tanh^n(x)-1\} \enskip ,
\end{equation}
which can be calculated analytically. After choosing some power sets $\{\mathbf{p}_s\}$, the parameters $f_s^{(n)}$ can be determined via fitting the expression at such $(\Lambda_{i_1}^{I_1},\Lambda_{i_2}^{I_2},\dots,\Lambda_{i_N}^{I_N})$ points, for which the value of $\| \prod_{k}^n g_{i_k}^{I_k} -1 \|$ is known.
Considering that \textit{the critical behavior of Eq. (\ref{eq:gamma_simp}) must be independent from the power sets used in Eq. (\ref{eq:tanh_fit})} it is clear that all $x_I$'s must be equal. [Otherwise, arbitrary powers of $\epsilon$ would emerge in Eq. (\ref{eq:appb_gamma})]. 

Considering only the leading order, the simplest form of $\|g_i^1 g_j^1-1\|$ that couples $\lambda_i$ and $\lambda_j$ comes from the power sets $\mathbf{p}=(0,1)$ and $(1/2,1/2)$:
\begin{equation}
\label{eq:profile2_base}
\|g_i^1g_j^1-1\| \approx \frac{f_1^{(2)}}{2}(\Lambda_i^1 + \Lambda_j^1) + f_2^{(2)} \sqrt{\Lambda_i^1 \Lambda_j^1} \enskip .
\end{equation}
Since $f_1^{(2)}+f_2^{(2)}=\int_{-\infty}^\infty dx \left\{ \tanh^2(x)-1 \right\}=-2$, Eq. (\ref{eq:profile2_base}) reduces to
\begin{equation}
\label{eq:profile2_2}
\|g_i^1g_j^1-1\| \approx H_2 (\Lambda_i^1+\Lambda_j^1) -2(1+H_2)\sqrt{\Lambda_i^1 \Lambda_j^1} \enskip ,
\end{equation}
where $H_2$ can be determined by solving
\begin{equation}
I_2[\eta]= H_2 \left( 1+ \frac{1}{\eta} \right) - \frac{2(1+H_2)}{\sqrt{\eta}}
\end{equation}
for a chosen ratio $\eta=\Lambda_i^1/\Lambda_j^1=\lambda_i^1/\lambda_j^1 \neq 1$, where $I_2[\eta]=\int_{-\infty}^{+\infty}dx\{\tanh(x)\tanh(\eta \cdot x)-1\}$. The same derivation applies for $\|(g_i^1 g_j^1)^2-1\|$, yielding
\begin{equation}
\label{eq:profile2_22}
 \| (g_i^1 g_j^1)^2 -1 \| \approx H_4 (\Lambda_i^1+\Lambda_j^1) -2 \left(\frac{4}{3}+H_4 \right) \sqrt{\Lambda_i^1 \Lambda_j^1} \enskip ,
\end{equation}
where $H_4$ can be determined via
\begin{equation}
I_4[\eta]= H_4 \left( 1+ \frac{1}{\eta} \right) - \frac{2(4/3+H_4)}{\sqrt{\eta}} \enskip ,
\end{equation}
where $I_4[\eta]=\int_{-\infty}^{+\infty} dx \{ [\tanh(x)\tanh(\eta \cdot x)]^2-1\}$. Using Eq. (\ref{eq:profile2_22}) a reasonable approximation of $\|g_i^1g_j^1g_k^1g_l^1-1\|$ reads as:
\begin{equation}
\label{eq:profile2_4}
\begin{split}
\|g_i^1 g_j^1 g_k^1 g_l^1-1\| \approx &  \left( \frac{3 H_4}{4} + \frac{1}{3} \right) (\Lambda_i^1+\Lambda_j^1 + \Lambda_k^1 + \Lambda_l^1) - \\
& - \left( \frac{2}{3}+\frac{H_4}{2} \right) \sum_{m,n} \sqrt{\Lambda_m^1 \Lambda_n^1} \enskip .
 \end{split} 
\end{equation}
Note that Eq. (\ref{eq:profile2_4}) reduces to $\|(g_i^1 g_j^1)^2-1\|$ in case of two equal pairs in $\{\lambda_i,\lambda_j,\lambda_k,\lambda_l\}$.

Now we can evaluate the anisotropic interfacial free energy as follows: First we calculate Eq. (\ref{eq:gamma_simp}) divided by Eq. (\ref{eq:isogamma}):
\begin{equation}
\label{eq:gamma_simp2}
\begin{split}
& \frac{\gamma(\mathbf{n})}{\gamma} = \sum_i \left[ -\frac{3}{N_1} \tilde{\lambda}_i + \frac{\zeta(\mathbf{n}\cdot\mathbf{\Gamma}_i^1,\Gamma_1)}{2 C N_1} \frac{1}{\tilde{\lambda}_i} \right] + \\
& \frac{3}{N_3} \sum_{i,j,k} \left[ \sum_{m,n} f_2(\tilde{\lambda}_m,\tilde{\lambda}_n)\right] \delta_{i,j,k}^{1,1,1} - \\
&  \frac{1}{N_4} \sum_{i,j,k,l} \left[ f_4(\tilde{\lambda}_i,\tilde{\lambda}_j,\tilde{\lambda}_k,\tilde{\lambda}_l) + \frac{3}{4} \sum_{m,n} f_2(\tilde{\lambda}_m,\tilde{\lambda}_n) \right] \delta_{i,j,k,l}^{1,1,1,1} \enskip ,
\end{split}
\end{equation}
where $\tilde{\lambda}_i=\lambda_i^1/\lambda_1$ (the interface thickness relative to the isotropic solution $\Lambda_1=\lambda_1/\sqrt{\epsilon}$), whereas
\begin{eqnarray}
\label{eq:def_f2}
f_2(\tilde{\lambda}_i,\tilde{\lambda}_j) &=& \| g_i^1 g_j^1 -1\|/(-2\Lambda_1)\\
\label{eq:def_f4}
f_4(\tilde{\lambda}_i,\tilde{\lambda}_j,\tilde{\lambda}_k,\tilde{\lambda}_l) &=& \|g_i^1g_j^1g_k^1g_l^1-1\| / [(-8/3)\Lambda_1] \enskip .
\end{eqnarray}

\section*{APPENDIX E: DETERMINING THE GINZBURG-LANDAU GRADIENT MATRICES}

First we modify Eq. (\ref{eq:gamma_aniso_gl}) by subtracting $(1/2) \Delta f[\vec{\phi}_S]$ from the integrand in order to achieve finite surface contributions. Note that $\Delta f[\vec{\phi}_S]\equiv 0$, therefore this modification has no effect on Eq. (\ref{eq:gamma_aniso_gl}). 

Next, we assume that the planar equilibrium solution read as:
\begin{equation}
\label{eq:gl_interface}
\phi^*_i(\xi) = \frac{1+g_i(\xi)}{2} \enskip .
\end{equation} 
Using Eq. (\ref{eq:gl_interface}) and Eq. (\ref{eq:anisoGL}) in Eq. (\ref{eq:gamma_aniso_gl}) results in
\begin{equation}
\label{eq:amplitude_aniso_gamma_gl}
\begin{split}
\gamma_{GL}(\mathbf{n}) = & \frac{\kappa}{4} \sum_{i,j} s_{ij}\frac{\| \partial g_i \partial g_j \|}{\Lambda_i \Lambda_j} -\frac{w}{2 N_1} \sum_i \Lambda_i - \\
& - \frac{w}{4 N_3} \sum_{i,j,k} \left[ \sum_{m,n}^{i,j,k} \| g_m^M g_n^N -1 \|\right] \delta^{1,1,1}_{i,j,k} + \\
& + \frac{w}{16 N_4} \sum_{i,j,k,l} \left[ \| g_i^I g_j^J g_k^K g_l^L -1 \| + \right. \\
& \left. +  \sum_{m,n}\| g_m^M g_n^N -1 \| \right] \delta^{1,1,1,1}_{i,j,k,l} \enskip .
\end{split}
\end{equation} 
In addition, taking the leading order of Eq. (\ref{eq:gamma_simp}), substituting $A_1=a_1\sqrt{\epsilon}$, $\bar{\psi}=c_\psi\sqrt{\epsilon}$ with Eqns. (\ref{eq:coefficient1}) and (\ref{eq:coefficient2}), and considering Eqns. (\ref{eq:amplitudew}) and (\ref{eq:amplitudekappa}) yields
\begin{equation}
\label{eq:amplitude_aniso_gamma_base}
\begin{split}
\gamma_{PFC}(\mathbf{n}) = & \sum_{i} \frac{\kappa }{C N_1} \frac{\zeta(\mathbf{n}\cdot\mathbf{\Gamma}_i^1,\Gamma_1)}{3\Lambda_i} -   \frac{w}{2 N_1}\sum_i \Lambda_i  + \\
& - \frac{w}{4 N_3} \sum_{i,j,k} \left[   \sum_{m,n}^{i,j,k} \| g_m^M g_n^N -1\| \right] \delta^{1,1,1}_{i,j,k} + \\
& + \frac{w}{16 N_4} \sum_{i,j,k,l} \left[  \| g_i^I g_j^J g_k^K g_l^L -1 \|  + \right. \\
& \left .+  \sum_{m,n} \| g_m^M g_n^N -1 \|   \right] \delta^{1,1,1,1}_{i,j,k,l} \enskip .
\end{split}
\end{equation}
Comparing Eq. (\ref{eq:amplitude_aniso_gamma_base}) and  (\ref{eq:amplitude_aniso_gamma_gl}) indicates that $\mathbb{S}$ must be diagonal, namely, $s_{ij}=s_i \delta_{ij}$, and $s_i=\zeta(\mathbf{n}\cdot \mathbf{\Gamma}_i^1)/(N_1 C)$. Note that $\sum_i s_{ii} \equiv 1$.

The corresponding coefficient matrices $\mathbb{A}_i$ in Eq. (\ref{eq:amplitude_aniso}) can be determined as follows. First we express $\nabla\phi_i$ in an Euclidean coordinate system where the $x$ direction is parallel to the interface normal. Characterizing $\mathbf{n}$ by the $(\alpha,\beta,\gamma)$ Euler-angles yields the transformation matrix $\mathbb{R}=\mathbb{R}^z_\alpha\mathbb{R}^y_\beta\mathbb{R}_\gamma^x$, where $\mathbb{R}_\delta^w$ denotes a 3D rotation matrix by angle $\delta$ around axis $w$ in the original coordinate system. The gradient term can be then expressed as:
\begin{equation}
\label{eq:appglnabla}
\sum_i \nabla'\phi_i \mathbb{M}_i \nabla'\phi_i \enskip ,
\end{equation} 
where $\mathbb{M}_i=(\mathbb{R}^T \cdot  \mathbb{A}_i \cdot \mathbb{R})$. For the planar equilibrium interface of normal $\mathbf{n}$ Eq. (\ref{eq:appglnabla}) reduces to 
\begin{equation}
\sum_i m^{(i)}_{11} (\partial_\xi \phi_i)^2 \enskip ,
\end{equation} 
where $m_{11}^{(i)} = \mathbf{n}^T \cdot \mathbb{A}_i \cdot \mathbf{n}$. Considering Eq. (\ref{eq:gamma_aniso_gl}), $m_{11}^{(i)}\equiv s_i=[\zeta_0 + \zeta_1 \cdot (\mathbf{n}\cdot \mathbf{\Gamma}_i^1)^2]/(N_1 C)$ yields
\begin{equation}
\mathbb{A}_i = \frac{1}{N_1 C} \left[\zeta_0 \cdot \mathbb{I} + \zeta_1 \cdot (\mathbf{\Gamma}_i^1 \otimes \mathbf{\Gamma}_i^1) \right] \enskip .
\end{equation}

\end{document}